\documentclass[usenatbib]{mn2e}
\usepackage{epsfig,aastex_hack, amssymb}

\newcommand{\mincir}{\raise
-2.truept\hbox{\rlap{\hbox{$\sim$}}\raise5.truept 
\hbox{$<$}\ }}
\newcommand{\magcir}{\raise
-2.truept\hbox{\rlap{\hbox{$\sim$}}\raise5.truept
\hbox{$>$}\ }}
\newcommand{\minmag}{\raise-2.truept\hbox{\rlap{\hbox{$<$}}\raise
6.truept\hbox
{$>$}\ }}

\newcommand{\beq}{\begin{equation}}
\newcommand{\eeq}{\end{equation}}

\newcommand{\be}{\begin{equation}}
\newcommand{\ee}{\end{equation}}
\newcommand{\ba}{\begin{eqnarray}}
\newcommand{\ea}{\end{eqnarray}}
\newcommand{\brr}{\begin{array}}
 
\newcommand{\err}{\end{array}}
\newcommand{\bc}{\begin{center}}
\newcommand{\ec}{\end{center}}

\newif\ifAMStwofonts
\AMStwofontstrue

\def\lsim{\mathrel{\hbox{\rlap{\hbox{\lower4pt\hbox{$\sim$}}}\hbox{$<$}}}}
\def\gsim{\mathrel{\hbox{\rlap{\hbox{\lower4pt\hbox{$\sim$}}}\hbox{$>$}}}}

\def\bullname{1E 0657-56 }

\def\LCDM{$\Lambda$CDM }

\newcommand{\kms}{\>{\rm km}\,{\rm s}^{-1}}

\newcommand{\kpch}{\>h^{-1} {\rm kpc}}

\newcommand{\Mpch}{\>h^{-1} {\rm Mpc}}
\newcommand{\Mpc}{\>{\rm Mpc}}

\newcommand{\Msolh}{\>h^{-1} M_{\odot}}

\DeclareMathAlphabet{\mathsc}{OT1}{cmr}{m}{sc}
\def\testbx{bx}%
\DeclareRobustCommand{\ion}[2]{%
\relax\ifmmode
\ifx\testbx\f@series
{\mathbf{#1\,\mathsc{#2}}}\else
{\mathrm{#1\,\mathsc{#2}}}\fi
\else\textup{#1\,{\mdseries\textsc{#2}}}%
\fi}

\title[How Rare is the Bullet Cluster?]
{How Rare is the Bullet Cluster?}
\author[E. Hayashi and S.D.M. White] {Eric Hayashi
and Simon D.M. White\\Max Planck Institute for
Astrophysics,  Karl-Schwarzschild Strasse 1,  D-85748 Garching, Germany}

\begin{document}

\maketitle
\begin{abstract}
The galaxy cluster \bullname has a bullet-like subcluster that is moving away
from the centre of the main cluster at high speed.  \cite{MARKEVITCH04} recently
estimated a relative velocity of $V_{\rm bullet} = 4500^{+1100}_{-800}~\kms$,
based on observations of the bow shock in front of the subcluster.  The weak
lensing analysis of \cite{CLOWE04} indicates that a substantial secondary mass
peak is associated with this subcluster.  We estimate the likelihood of such a
configuration by examining the distribution of subhalo velocities for clusters
in the Millennium Run, a large \LCDM cosmological simulation.  We find that the
most massive subhalo has a velocity as high as that of the bullet subcluster in
only about 1 out of every 100 cluster-sized halos.  This estimate is strongly
dependent on the precise velocity adopted for the bullet.  One of the ten most
massive subhalos has such a high velocity about $40\%$ of the time.  We conclude
that the velocity of the bullet subcluster is not exceptionally high for a
cluster substructure, and can be accommodated within the currently favoured
\LCDM comogony.
\end{abstract}


\section{Introduction}
\label{sec:intro}

Galaxy cluster \bullname is a singular example of a merging system.  X-ray
observations of the cluster have revealed a bow shock propagating in front of a
bullet-like gas cloud moving away from the core of the main cluster.  Based on
the gas density jump across the shock front, \cite{MARKEVITCH02} first derived
an estimate of velocity of the bullet subcluster, $V_{\rm bullet}\simeq 4000
\kms$.  The unique geometry of the system was recently exploited by
\cite{CLOWE04}, who combined the X-ray observations with weak lensing analysis
to show that the mass distribution does not follow that of the hot cluster gas,
but could plausibly be associated with a collisionless dark matter component.
These authors argue that the data are not easily reconciled with modified
Newtonian Dynamics \citep[][MOND]{MILGROM83A} in which the mass budget would be
dominated by the gas.  \cite{MARKEVITCH04} also use \bullname to derive
constraints on the cross-section for self-interaction of the dark matter based on
the spatial offset between the peaks in the X-ray and mass distributions.

In this paper we estimate the probability of finding such a high velocity
subcluster in a \LCDM cosmology by examining the statistics of dark matter
substructure halos (subhalos) in a very large cosmological N-body simulation.
We calculate the distribution of subhalo velocities relative to their host halos
for a large sample of cluster-sized hosts and use this to determine the fraction
of clusters which contain a high velocity subcluster.

\section{Properties of the Bullet Cluster}
\label{sec:bulletprops}

The cluster \bullname was discovered as an extended source in the ${\it
Einstein}$ imaging proportional counter (IPC) database and was identified as a
rich cluster of galaxies by \cite{TUCKER95}.  \cite{TUCKER98} established the
cluster redshift as $z=0.296$ and identified the bullet subcomponent in a
$ROSAT$ X-ray image of the cluster.  They also measured the temperature of the
hot cluster gas as $kT \sim 17~ {\rm keV}$, making \bullname one of the hottest
known clusters.  \cite{MARKEVITCH02} subsequently revised this to an average
cluster temperature of 14-15 keV, albeit with large spatial variations, based
on $Chandra$ observations of the system.

The transverse separation between the bullet subcluster and the centre of the
main cluster is $\sim 0.48~\Mpch$ \citep{CLOWE04}. \cite{MARKEVITCH04} estimate
the velocity of the bullet subcluster, $V_{\rm bullet} = 4500^{+1100}_{-800}
\kms$; here and elsewhere we adopt a Hubble constant of $H_0=73~\kms/\Mpc$.  The
line-of-sight velocity of the subcluster is $\sim 600 \kms$ relative to the
cluster centre \citep{BARRENA02}, so the direction of the bullet's motion is
very nearly in the plane of the sky.

\cite{CLOWE04} use weak lensing to model the mass distribution of the main
cluster and the bullet subcluster.  They find that the main cluster is well fit
by a \citet[NFW]{nfw96} profile with concentration $c_{200}=3.0$ and virial
radius $r_{200}=1.64~\Mpch$, corresponding to a virial mass and velocity of
$M_{200}=2.16 \times 10^{15} \Msolh$ and $V_{200}=(G M_{200}/r_{200})^{1/2}=2380~
\kms$, respectively.  They also find a secondary peak in the mass distribution
that is clearly associated with the bullet subcluster, and they measure a mass
of $(5.3 \pm 1.5) \times 10^{13}~\Msolh$ for the subcluster within a cylinder of
radius $0.11~\Mpch$.  We adopt these values for the mass of the main cluster and
the bullet rather than estimates based on the cluster galaxy velocity dispersion
or the gas temperature since the latter rely on assumptions of isotropy and
hydrostatic equilibrium which may not be valid for a merging cluster system. We
note that the virial mass we adopt corresponds to an overall X-ray temperature
of $kT \simeq 15-16~\rm keV$ according to the mass-temperature relations of
\cite{ARNAUD05}, in good agreement with the observed X-ray temperature.

To summarize, the main properties of this system relevant to this study are as
follows:

\begin{itemize}
\item Cluster \bullname is a very massive cluster, as evidenced by its high
  temperature and confirmed by the high mass detected by weak lensing analysis.

\item The bullet subcluster is the most massive substructure in the cluster, and
  represents a mass $\sim 2\%$ that of the main cluster.

\item The bullet subcluster is moving away from the main cluster at velocity
  $V_{\rm bullet} \simeq 4500~\kms = 1.9~V_{200}$ and is at least $0.48~\Mpch
  \simeq 0.3~r_{200}$ from the cluster centre.  
\end{itemize}

In the following section we assess the likelihood of such an object by
 examining the statistics of dark matter halos in a large cosmological
 simulation.

\section{Searching for the Bullet}
\label{sec:bulletsearch}

This study makes use of the Millennium Run \citep{SPRINGEL05}, a very large
cosmological N-body simulation carried out by the Virgo
Consortium.\footnote{http://www.virgo.dur.ac.uk/}$^,$\footnote{http://www.mpa-garching.mpg.de/galform/virgo/millennium/}
In this simulation a flat \LCDM cosmology is adopted, with $\Omega_{\rm
dm}=0.205$ and $\Omega_b=0.045$ for the current densities in cold dark matter
and baryons, $h=0.73$ for the present dimensionless value of the Hubble
constant, $\sigma_8=0.9$ for the rms linear mass fluctuation in a sphere of
radius $8 \Mpch$ extrapolated to $z=0$, and $n=1$ for the slope of the
primordial fluctuation spectrum.  The simulation follows $2160^3$ dark matter
particles from $z=127$ to $z=0$ within a cubic region $500 \Mpch$ on a side.
The individual particle mass is thus $8.6 \times 10^8 \Msolh$, and the
gravitational force is softened with a Plummer-equivalent comoving softening of
$5 \kpch$. Initial conditions were generated using the Boltzmann code CMBFAST
\citep{SELJAK96} to generate a realization of the desired power spectrum which
was then imposed on a glass-like uniform particle load \citep{WHITE96}. A
modified version of the TREE-PM N-body code GADGET2
\citep{SPRINGEL01,SPRINGEL05} was used to carry out the simulation and full
particle data are stored at 64 output times approximately equally spaced in the
logarithm of the expansion factor.

In each output of the simulation, halos are identified using a
friends-of-friends (FoF) groupfinder with a linking length of $b=0.2$
\citep{DAVIS85}.  The virial radius, $r_{200}$, is defined for each FoF halo by
calculating the radius of a sphere, centered on the particle with the minimum
potential, that encompasses a mean density 200 times the critical value.  Each
FoF halo is decomposed into a collection of locally overdense, self-bound
substructures (or subhalos) using the SUBFIND algorithm of \cite{SPR01}.  Of
these subhalos, one is typically much larger than the others and contains most
of the mass of the halo.  We identify this as the main halo and subtract its
centre of mass velocity from that of the remaining subhalos to compute the
relative velocity between subhalos and their host halo.  Hereafter, we refer to
this as the subhalo velocity, not to be confused with the internal circular
velocity of the subhalo.

We search the Millennium simulation for halos whose most massive subhalo has a
velocity relative to the main halo comparable to that of the bullet subcluster,
i.e., $V_{\rm sub} > 1.9~V_{200}$.  We focus our search on cluster- and
group-sized halos in the $z=0.28$ output of the Millennium simulation, the
output closest in redshift to \bullname.  

\begin{figure}

\plotone{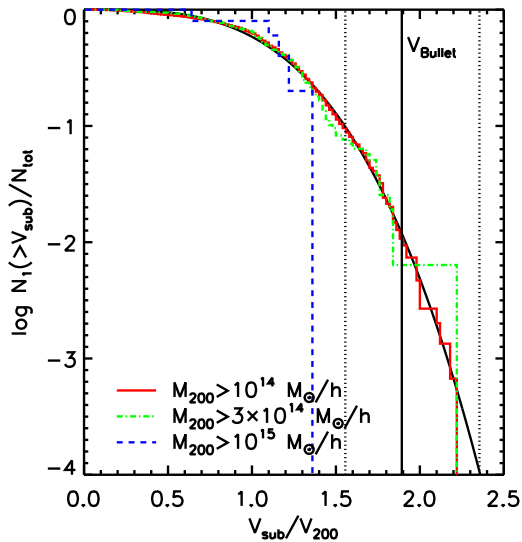}
\caption{Fraction of host halos whose most massive subhalo has velocity greater
  than $V_{\rm sub}/V_{200}$, for three different host halo mass ranges.  The
  number of halos with $M_{200} > 10^{14}~\Msolh$, $3~\times~10^{14}~\Msolh$,
  and $10^{15}~\Msolh$ is 1491, 157, and 5, respectively.  The $M_{200} >
  10^{14}~\Msolh$ distribution is well fit by a eq.~\ref{eq:fit}, shown as the
  solid curve.  The vertical lines indicate the velocity of the bullet
  subcluster (solid line) and lower and upper limits (dotted lines).  About
  $1\%$ of hosts have a most massive subhalo with velocity greater than that of the
  bullet.  This percentage increases (decreases) to $10\%$ ($\ll 0.1\%$) for the
  lower (upper) limits of the bullet velocity.}
\label{fig:vcumhist123}
\end{figure}

Figure~\ref{fig:vcumhist123} shows the cumulative distribution of velocities for
the most massive subhalo in host halos with $M_{200} > 10^{14}~\Msolh$,
$3~\times~10^{14}~\Msolh$, and $10^{15}~\Msolh$.  The number of hosts in each of
these mass ranges is $N_{\rm hosts} =1491$, 157, and 5, respectively.  We note
that \bullname is one of the hottest known clusters and that very few clusters
in the Millennium Run have masses comparable to that of \bullname due to the
limited volume of the simulation. According to the cluster temperature function
of \cite{HENRY04}, one expects les than 0.1 clusters as hot as \bullname in a
volume similar to that of the Millennium simulation.  In fact, at $z=0.28$, the
Millennium simulation contains one cluster halo with $M_{200} > 2 \times
10^{15}~\Msolh$.

The shape of the velocity distribution appears relatively insensitive to host
 halo mass and the median subhalo velocity is $1.1~V_{200}$.  We note that all
 subhalos have velocities have velocities much less than the maximum escape
 velocity, $v_{\rm esc} \simeq 3.3~V{200}$ for an NFW potential with $c_{200}
 \simeq 6$.

The fraction of halos with $M_{200} > 10^{14}~\Msolh$ whose most massive subhalo
has a velocity greater than that of the bullet subcluster (scaled to the virial
velocity of the host halo) is 16 out of 1491 or approximately $1\%$.  However
the velocity distribution drops steeply at high velocities, and this percentage
increases (decreases) to $10\%$ ($\ll0.1\%$) if the lower (upper) limit is adopted
for the bullet velocity.  We note that this is in agreement with the simulation
results of \cite{GILL05} who also find a small but significant fraction of high
velocity subhalos in cluster-sized host halos.

We find that the velocity distribution is well fit by a function of the
following form, shown as the solid curve in Figure~\ref{fig:vcumhist123}:

\begin{equation}
\log \frac{N_1(>V_{\rm sub})}{N_{\rm hosts}} = -\left (\frac{V_{\rm
    sub}/V_{200}}{v_{10\%}}\right)^\alpha, 
\label{eq:fit}
\end{equation}

where $v_{10\%}$ is the velocity in units of $V_{200}$ at which the fraction of
halos drops to $0.1$.  Fitting this function to the $M_{200} > 10^{14}~\Msolh$
distribution at $z=0.28$ yields best fit values of $v_{10\%}=1.55$ and
$\alpha=3.3$.  We note that the best fit value of $v_{10\%}$ tends to decrease
with decreasing redshift: $v_{10\%}=1.64, 1.52,$ and $1.32$ at $z=0.5, 0.11,$
and $0$, respectively. The value of $\alpha$ shows no significant change with
redshift and is typically consistent with $\alpha \simeq 3.0-3.1$.  We attribute
this to the increase in the virial velocity of the halo with respect to the
velocities of subhalos at the time of infall.  Indeed, the mean $V_{200}$
increases from $780~\kms$ at $z=0.5$ to $920~\kms$ at $z=0.0$, whereas the mean
subhalo velocity decreases by only a few percent, from $1190~\kms$ to
$1129~\kms$, over the same period.

Having quantified the likelihood that a halo has a most massive subhalo with a
velocity comparable to that of the bullet subcluster, we now estimate the
probability of finding a high velocity subhalo amongst a halo's $n$ most massive
subhalos. The probability of drawing at least one subhalo with velocity $>V_{\rm
sub}$ from the $n$ most massive subhalos is given by

\begin{equation}
P(>V_{\rm sub}) = 1-\prod_{i=1}^n \left(1-\frac{N_i(>V_{\rm sub})}{N_{\rm hosts}}\right).
\label{eq:probprod}
\end{equation}

\begin{figure}

\plotone{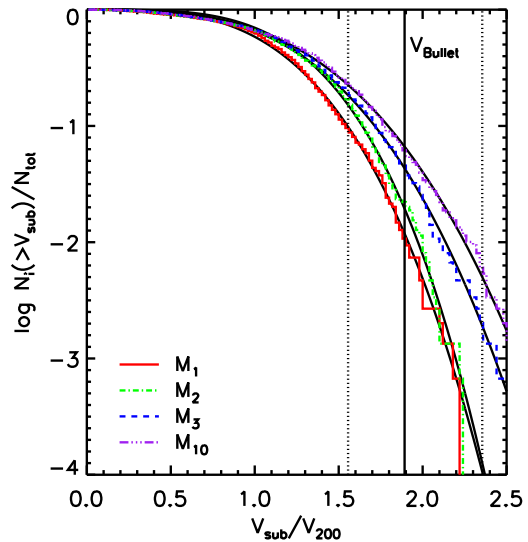}
\caption{Fraction of host halos whose $i$-th most massive subhalo has velocity
  greater than $V_{\rm sub}/V_{200}$, for 1st, 2nd, 3rd and 10th most massive
  subhalos.  More massive subhalos are biased toward lower velocities. Solid
  curves show fits with eq.~\ref{eq:fit}, with values for $v_{10\%}$ of $ 1.55,
  1.64, 1.72,$ and $1.79$ in order of increasing subhalo rank, respectively.}
\label{fig:vcumhist129}
\end{figure}

We investigate whether subhalo velocity is independent of subhalo mass in
Figure~\ref{fig:vcumhist129} by comparing the velocity distributions for the
1st, 2nd, 3rd and 10th most massive subhalos in halos of mass $M_{200} >
10^{14}~\Msolh$.  We find that more massive subhalos are slightly biased toward
lower velocities.  The best fit values of $v_{10\%}$ are $1.55$, $1.64$, $1.71$,
and $1.79$ for the velocity distributions of the 1st, 2nd, 3rd and 10th most
massive subhalos, and $v_{10\%} \simeq 1.8$ for subhalos of higher rank.  We
find that the following formula accurately describes the trend:

\begin{equation}
v_{10\%}(i)=1.8 - 0.25~\exp(-0.45~(i-1)).
\label{eq:v10fit}
\end{equation}

Note that the value of $\alpha$ decreases slightly with increasing subhalo rank,
but we find that a constant value of $\alpha=3.1$ provides an adequate fit to
the distributions for all subhalo ranks.

We combine eqs.~\ref{eq:fit}, \ref{eq:probprod}, and \ref{eq:v10fit} in order
to predict the probability of at least one subhalo with $V_{\rm sub} >
1.9~V_{200}$.  This gives a probability of $39.7\%$ which agrees well with the
actual fraction, $40.8\%$, found for halos in the Millennium simulation with
$M_{200} > 10^{14}~\Msolh$.  In comparison, if we adopt a constant value for
$v_{10\%}$ of $1.55~(1.8)$ we predict a probability of $12.4\%~(49.3\%)$.

In order to convert the subhalo mass rank into a fraction of the host halo mass,
we compute the subhalo mass function for our halo sample.  We find that the
differential subhalo mass function for halos with $M_{200} > 10^{14}~\Msolh$ is
well fit by a power law with slope $-0.88$, in agreement with the mass functions
of cluster halos presented by \cite{DELUCIA04}.  Integrating the differential
mass function yields the total number of subhalos with mass greater than $M_{\rm
sub}$:

\begin{equation}
\log N(>M_{\rm sub}) =  -0.88 \log \frac{M_{\rm sub}}{M_{200}} -1.67.
\label{eq:massfncum}
\end{equation}

Solving this equation for $M_{\rm sub}$ gives the subhalo mass corresponding to
subhalo mass rank $N$.  For example, for $N=10$ we find $M_{\rm sub} =
0.001~M_{200}$, corresponding to subhalos of mass $10^{12}~\Msolh$ for
cluster-sized host halos of mass $M_{200} \simeq 10^{15}~\Msolh$.  Our previous
result therefore implies that $40\%$ of massive cluster halos have at least one
subhalo of mass greater than $10^{12}~\Msolh$ with a velocity comparable to that
of the bullet subcluster.

We now return to the bias in the velocity distribution of the most massive
subhalos (see Figure~\ref{fig:vcumhist123}).  This is related to the fact that more
massive substructures are preferentially located in the external regions of
their host halos, as noted by \cite{DELUCIA04}.  These subhalos are closer to
the apocentre of their orbits, and therefore have lower velocities compared to
subhalos near pericentre.  Conversely, we expect to find high velocity subhalos
near the centre of their host halos.  Indeed, of the 16 high velocity subhalos
that are bullet subcluster candidates, all but one are located at $r <
0.6~r_{200}$, whereas only $15\%$ of total sample of most massive subhalos are
found within this radius.  This correlation between subhalo velocity and
clustercentric distance was also noted by \cite{GILL05}.

\begin{figure}
\plotone{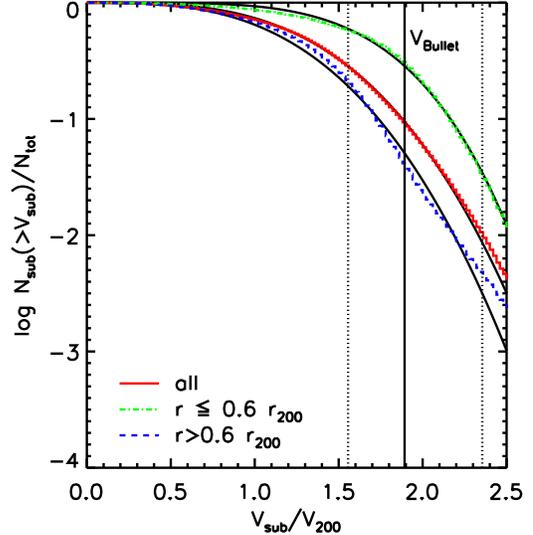}
\caption{Fraction of subhalos with velocity greater than $V_{\rm sub}/V_{200}$,
  at different radii.  Subhalos in the outer regions are biased toward lower
  velocities. Solid curves show fits with eq.~\ref{eq:fit}, with values for
  $v_{10\%}$ of $1.74, 2.16$ and $1.87$ for subhalos with $r > 0.6~r_{200}$,
  with $r < 0.6~r_{200}$ and for the total sample, respectively.}
\label{fig:vcumhistrlt}
\end{figure}

In Figure~\ref{fig:vcumhistrlt} we compare the velocity distributions of
subhalos in the inner and outer regions.  Subhalos within the central
$0.6~r_{200}$ of the host halo (about half of the total within $r_{200}$) are
indeed biased to higher velocities.  The best fit values of $v_{10\%}$ are 1.74,
2.16, and 1.87 for the velocity distributions of subhalos with $r \le
0.6~r_{200}$, $r > 0.6~r_{200}$ and the combined sample, respectively.  We also
note that velocity distribution of outer subhalos is not as well fit by
eq.~\ref{eq:fit} as the other distributions, however the deviations are
typically $\lesssim 0.1$ dex.  

Finally, we note that the direction of the bullet subcluster's velocity is an
additional constraint that we have not so far considered.  As stated in
\S\ref{sec:bulletprops}, the bullet subcluster is moving {\it away} from the
cluster centre, i.e., has a positive radial velocity.  Of the 16 bullet
subcluster candidates, five have positive radial velocities.  Overall, we find
that a smaller fraction of massive subhalos tend to have positive radial
velocities compared to less massive subhalos; this fraction is $20\%$, $30\%$
and $40\%$ for the most massive, second most massive, and tenth most massive
subhalo samples.

We attribute this to the mass loss and disruption of subhalos by tidal
stripping, which occurs near orbital pericentre.  This can decrease the mass of
a subhalo by as much as $90\%$ \citep[e.g.,][]{HAYASHI03}, potentially
downgrading the mass rank of more massive subhalos after they pass through
pericentre.  Less massive subhalos can be disrupted altogether resulting in the
depletion of subhalos with positive radial velocities.  However, the velocity
distributions of subhalos with positive and negative radial velocities are very
similar; subhalos with positive radial velocities are slightly biased toward
lower velocities, but the difference in the best fit values of $v_{10\%}$ and
$\alpha$ are typically $\sim0.05$ and $\sim0.5$, respectively.  We therefore
conclude that if the direction of the bullet subcluster is considered, the
probabilities we have estimated are reduced by about $60-70\%$, the fraction of
subhalos with negative radial velocities.

\section{Conclusions}
\label{sec:conc}

The bullet subcluster is a massive substructure in galaxy cluster \bullname
moving at a high velocity relative to the centre of the main cluster.  In units
of the virial velocity of the cluster, the velocity of the bullet subcluster is
$V_{\rm bullet}\simeq 1.9~V_{200}$.  We have examined the distribution of subhalo
velocities relative to their host halos in a large cosmological simulation in
order to assess the likelihood of a system like 1E 0657-56.

We calculate the velocity distribution of the most massive subhalos in 1491 host
halos with virial masses $M_{200} > 10^{14}~\Msolh$ and find that about 1 in 100
have velocities comparable to that of the bullet subcluster if the best estimate
of \cite{MARKEVITCH04} is adopted for the bullet velocity.  However, this
fraction depends strongly on the velocity of the bullet and ranges from
$\ll0.1\%$ to $10\%$ for the upper and lower limits on the bullet velocity,
respectively.  We find that more massive subhalos are biased towards lower
velocities, as are subhalos in the outer regions of halos.  Taking this into
account, we find that at least one of the ten most massive subhalos has a
velocity as high as that of the bullet subcluster in $40\%$ of all host halos.
We also find that subhalos are preferentially found to be moving toward the
centres of halos, most likely a result of tidal depletion of subhalos at
pericentre.  With this additional constraint, the likelihood of the bullet
cluster drops to about 1 in 500.  We conclude that the best estimate for the
velocity of the bullet subcluster is high but not extraordinary considering the
mass of its host cluster.  It is a rare but not an impossible event within the
currently favoured \LCDM comogony.

\section*{Acknowledgments}
\label{acknowledgements}
This work was conceived after discussions with M.~Brada\v{c}.  We thank
I.~McCarthy, M~.Balogh, and S.~McQueen for useful discussions.  

\bibliographystyle{astron}
\bibliography{stan}

\end{document}